\documentclass[pra,aps,twocolumn,10pt,showpacs,nofootinbib]{revtex4-1}
\usepackage[utf8]{inputenc}
\usepackage[T1]{fontenc}
\usepackage{amsmath,amsthm}
\usepackage{amsfonts}
\usepackage{amssymb}
\usepackage{mathcmd}
\usepackage{graphicx} 

\usepackage{braket}
\usepackage{bm}
\usepackage{mathrsfs}
\usepackage{stmaryrd}
\usepackage{esint}
\usepackage{xcolor}

\usepackage{tikz}
\usepackage{rotating} 
\usepackage{float}

\newtheorem*{definition}{Definition}

\newcommand{\mi}{\mathrm{i}}
\newcommand{\me}{\mathrm{e}}
\newcommand{\md}{\mathrm{d}}

\newcommand{\Nt}{N_{\mathrm{t}}}

\begin{document}

\title{Particle-Number Threshold for Non-Abelian Geometric Phases}
\author{Julien Pinske}
\author{Vincent Burgtorf}
\author{Stefan Scheel}
\email{stefan.scheel@uni-rostock.de}
\affiliation{Institut f\"ur Physik, Universit\"at Rostock, 
Albert-Einstein--Stra{\ss}e 23-24, D-18059 Rostock, Germany}

\date{\today}

    \begin{abstract}
    When a quantum state traverses a path, while being under the influence of a gauge potential, it acquires a geometric phase that is often more than just a scalar quantity. 
    The variety of unitary transformations that can be realised by this form of parallel transport depends crucially on the number of particles involved in the evolution.
    Here, we introduce a particle-number threshold (PNT) that assesses a system's capabilities to perform purely geometric manipulations of quantum states.
    This threshold gives the minimal number of particles necessary to fully exploit a system's potential to generate non-Abelian geometric phases. 
    Therefore, the PNT might be useful for evaluating the resource demands of a holonomic quantum computer. 
    We benchmark our findings on bosonic systems relevant to linear and nonlinear quantum optics.
    \end{abstract}
    
    \maketitle

    \section{Introduction}
    \label{sec:Intro}
    
    The evolution of a quantum state, in the presence of some potential, is completely determined by Schrödinger's equation which incorporates aspects such as the system's spectrum, or the overall evolution time. 
    If the system undergoes slow (adiabatic) changes, the evolving state remains unaffected by these dynamical contributions (i.e., the dynamical phase factors out).
    Instead, its wave function acquires a phase factor that only depends on the geometry of the path the quantum state has traversed. 
    This was first noticed by Berry \cite{B84} who pointed out that, unlike dynamical phases, a geometric phase cannot be removed by a rescaling of the energy (gauge transformation). 
    A famous example for this is the Aharonov-Bohm effect \cite{AB59}, in which the wave function of an electron traveling around a solenoidal magnetic field picks up a phase proportional to the magnetic flux through the surface enclosed by the trajectory of the electron. 
    Pancharatnam studied the phenomenon in the context of classical optics \cite{P56}, where it manifests itself in states of polarisation.
    It was pointed out by Simon \cite{S83} that this purely geometric signature of a quantum evolution has to be attributed to parallel transport of the state vector along a path in a (projective) Hilbert space. 

    If a quantum system supports a $d$-fold degenerate subspace $\mathscr{H}_0$ with eigenstates $\ket{\psi_a}$ ($a=1,\dots, d$), an initially prepared wave packet generically evolves into a superposition of the $\ket{\psi_a}$ when undergoing adiabatic changes, that is without population transfer to states of different energy \cite{BF28}. 
    Wilczek and Zee \cite{WZ84} associated such degeneracy of the spectrum with the possibility of emerging non-Abelian (i.e., noncommuting) gauge potentials. 
    In this case, the state after a time period $T$ does not only acquire a (scalar) geometric phase but differs from the initial one by a unitary $d\times d$ matrix. 
    
    If the Hamiltonian of the system is expressed through a set of physically accessible parameters $\{\kappa_\mu\}_{\mu=1}^M$ that change cyclically, i.e., $\kappa_\mu(0)=\kappa_\mu(T)$, the time evolution is associated with a closed path 
    $\gamma$ in the $M$-dimensional parameter space $\mathscr{M}$.
    The time evolution then takes the form of a quantum holonomy (non-Abelian geometric phase) \cite{WZ84}
    \begin{equation}
    \label{eq:Hol}
    U_A(\gamma)=\boldsymbol{\hat{\mathrm{P}}}\mathrm{exp}\Big(\oint_\gamma A \Big),
    \end{equation}
    where $A=\sum_{\mu=1}^M A_\mu\md\kappa_\mu$ is the adiabatic connection (non-Abelian gauge potential).
    Depending on the physical platform, the $\{\kappa_\mu\}_{\mu=1}^M$ might include external driving fields, subsystem couplings, or hopping probabilities between different states. 
    Due to the generally noncommuting nature of the connection, i.e., $[A_\mu,A_\nu]\neq 0$, the integration in Eq. \eqref{eq:Hol} has to be performed with respect to the path ordering $\boldsymbol{\hat{\mathrm{P}}}$. 
    The matrix-valued components of $A$ can be directly calculated from the eigenstates of the system, i.e.,
    \begin{equation}
    \label{eq:con}
    (A_\mu)_{ab}=\bra{\psi_b}\partial_\mu \ket{\psi_a},\quad \partial_\mu=\partial/\partial\kappa_\mu.
    \end{equation}
    By traversing different loops in $\mathscr{M}$ one can potentially access a variety of different unitaries $U_A(\gamma)$. 
    The set of all such transformations spans the holonomy group $\mathrm{Hol}(A)$. 
    It is a subset of the unitary group $\mathrm{U}(d)$. 
    In addition to their frequent occurrence in lattice-gauge theory \cite{BC20} and loop-quantum gravity \cite{R08}, holonomy groups turn out to be a crucial ingredient for geometric \cite{ZR99,PZ99} and topological \cite{NS08} quantum computation, where they constitute the fundamental gate set from which quantum algorithms are to be implemented. 
    
    The question of how many different unitaries can be harnessed by driving loops through $\mathscr{M}$ is therefore closely related to computational universality \cite{L95}, which holds if $\mathrm{Hol}(A)=\mathrm{U}(d)$.
    Not only does this require a $d$-fold degenerate subspace, but a large parameter space as well \cite{ZR99}.
    More recently, it was observed that the number of particles prepared in the subspace $\mathscr{H}_0$ might drastically alter the form of the holonomy $U_A(\gamma)$ \cite{BR12,PT20,NP22}.
    This is because the corresponding eigenstates $\ket{\psi_a}$ can differ in their particle number.
    In this work, quantum holonomies are studied in relation to the number of particles involved in the evolution.
    In the following, this issue is motivated through an illustrative example.

    \subsection{$\Lambda$-scheme of bosonic modes}
    \label{ssec:lambda}

\begin{figure}[h]
 	      \begin{tikzpicture}
 		\node at (0,0) {\includegraphics[width=7.5cm]{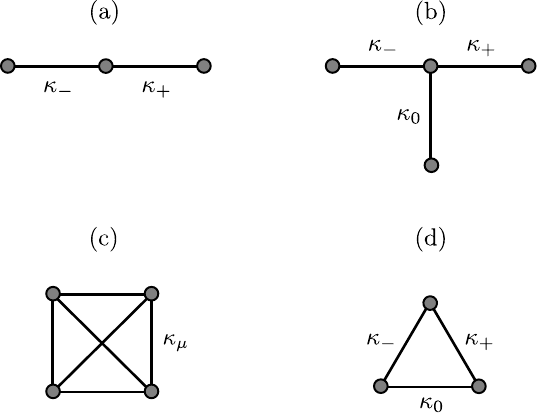}}; 
 	      \end{tikzpicture}
 	\caption{\label{fig:Mods} Graph representation of (bilinear) Hamiltonians, in which particle number exchange between the modes (vertices) is resembled by a connecting edge. 
    (a)  Schematic representation of three planarly arranged bosonic modes experiencing complex next-neighbour coupling $\kappa_{\pm}$. (b) Term scheme of the bosonic tripod structure in which the mode $\mathrm{c}$ exclusively couples to the outer modes $\mu=\pm,0$ via $\kappa_\mu$. (c) A four-mode fully connected graph, where each side can experience a different coupling $\kappa_\mu$. (d) Triangular graph of modes with coupling $\kappa_\mu$, $\mu=\pm,0$.}
    \end{figure}

    Consider a chain of three bosonic modes [Fig. \ref{fig:Mods} (a)].
    The outer modes $\hat{a}_\pm$ experience complex next-neighbour couplings $\kappa_{\pm}$ to the central mode $\hat{a}_{\mathrm{c}}$. 
    The Hamiltonian of the system reads
    \begin{equation}
    \label{eq:H1}
        \hat{H}=\kappa_{+}\hat{a}_+\hat{a}_{\mathrm{c}}^\dagger+ \kappa_{-}\hat{a}_{\mathrm{c}}\hat{a}_{-}^\dagger+\mathrm{H.c.}.
    \end{equation}
    Here, $\hat{a}_k^\dagger$ and $\hat{a}_k$ denote the bosonic creation and annihilation operators, respectively, and $\mathrm{H.c.}$ stands for the Hermitian conjugate.
    The Hamiltonian \eqref{eq:H1} is the bosonic counterpart of an atomic three-level system in $\Lambda$ configuration \cite{BT98}.
    Such systems are of practical interest as they describe linear-optical multiport systems \cite{GG19} and can be designed, for instance, in terms of integrated photonic waveguides \cite{SN10,LV07}. 

    Suppose a single photon is injected into one of the outer modes of the optical setup, with couplings $\kappa_\pm(t)$ varying slowly compared to the minimal energy gap $\sqrt{|\kappa_+|^2+|\kappa_-|^2}>0$ (level crossing neglected).
    In the adiabatic limit, the photon remains in the zero-eigenvalue eigenstate (aka dark state)
    \begin{equation*}
        \ket{D}=\sin\theta \ket{1_{+}}-\cos\theta\me^{\mi\varphi}\ket{1_{-}},
    \end{equation*}
    where $\tan\theta=|\kappa_-|/|\kappa_+|$, $\varphi=\mathrm{arg}(\kappa_+)-\mathrm{arg}(\kappa_-)$, and $\ket{1_{\pm}}=\hat{a}_\pm^\dagger\ket{\bm{0}}$ with $\ket{\bm{0}}$ denoting the three-mode vacuum.
    Here, the connection $A_{\varphi}=\mi\cos^2\theta$ is Abelian (while $A_\theta=0$).
    After traversing a closed path $\gamma$ in the $(\theta,\varphi)$ plane, the output state $\ket{\Psi(T)}=\me^{\mi\phi(\gamma)}\ket{\Psi(0)}$ picks up a geometric phase 
    \begin{equation}
        \label{eq:GP}
        \phi(\gamma)=\iint_{\mathscr{D}}\sin(2\theta)\md\varphi\md\theta,
    \end{equation}
    which depends on the area $\mathscr{D}$ enclosed by the loop $\gamma$. 

    Interestingly, injecting a second (indistinguishable) photon into the setup, leads to the two dark states 
    \begin{equation*}
        \begin{split}
        \ket{D_1}&=\sin^2\theta\ket{2_+}-\sqrt{2}\sin\theta\cos\theta\me^{\mi\varphi}\ket{1_+1_-}\\
        &\quad+\cos^2\theta \me^{2\mi\varphi}\ket{2_-},\\
        \ket{D_2}&=\frac{1}{\sqrt{2}}\big(\sin^2\theta\ket{2_-}+\cos^2\theta \me^{-2\mi\varphi}\ket{2_+}-\ket{2_\mathrm{c}}\big)\\
        &\quad+2\sin\theta\cos\theta\me^{-\mi\varphi}\ket{1_+1_-}.\\
        \end{split}
    \end{equation*}
    Consequently, $A_\varphi$ is now a matrix-valued quantity. 
    Naively, one might expect that this enables the generation of non-Abelian holonomies. 
    However, a direct evaluation of the Eq. \eqref{eq:Hol} leads to 
    \begin{equation}
        \label{eq:2LH}
        U_A(\gamma)=\begin{bmatrix}
        \me^{2\mi\phi(\gamma)}&0\\
        0&\me^{-2\mi\phi(\gamma)}\\
        \end{bmatrix}.
    \end{equation}
    It is immediately clear from Eq. \eqref{eq:2LH} that the transformations $U_A(\gamma)$ and $U_A(\gamma^\prime)$, induced by two arbitrary loops $\gamma$ and $\gamma^\prime$ in $\mathscr{M}$, always commute. 
    Hence, even though degeneracy of the system would allow for the generation of non-Abelian transformations, the actual holonomy group is still Abelian. 
    This phenomenon remains present when subjecting even more photons to the system \cite{PS22}; that is, while degeneracy scales up, the resulting holonomies are always commuting.

    The phenomenon that a system's degeneracy increases under the exposure to multiple photons is by no means a property unique to the Hamiltonian \eqref{eq:H1}. 
    Adding an additional mode to the $\Lambda$-scheme
    leads to a tripod structure [Fig. \ref{fig:Mods} (b)] that allows for any $\mathrm{U}(2)$ transformation between its single-photon dark states \cite{KT19,YZ22,SS22}. 
    Considering two photons, the dark subspace becomes four-dimensional. 
    However, as it was noticed in Refs. \cite{HN15,PT20} not all elements of the group $\mathrm{U}(4)$ can be designed in that way (one of the eigenstates decouples).
    Only, recently were these two-particle dynamics verified experimentally \cite{NP22}.   

    \subsection{Aim of the article}
    This simple introductory example hints at a more general question. 
    What is the number of particles $N$ injected into a given setup in order to generate the most versatile set of quantum holonomies? 
    After reviewing properties of the holonomy group in Sec. \ref{sec:curv}, we address this issue by introducing the particle-number threshold (PNT) in Sec. \ref{sec:PNT}. 
    The PNT of a quantum system gives the minimal number of particles necessary to fully exploit the system's potential for designing non-Abelian holonomies. 
    We discuss the basic properties of PNTs and present a number of different examples relevant to linear and nonlinear quantum optics.
    Finally, Sec.~\ref{sec:FIN} is reserved for a summary of the article as well as some concluding remarks.
    
    \section{Curvature and Universality}
    \label{sec:curv}
    
    If the composition of loops in $\mathscr{M}$ allows for the generation of any unitary on the $l$th eigenspace $\mathscr{H}_l$ of a Hamiltonian $\hat{H}$, the connection $A_l$ is said to be irreducible, and the holonomy group
    \begin{equation*}
        \mathrm{Hol}(A_l)=\big\{U_{A_l}(\gamma)\,|\,\gamma(0)=\gamma(T)\big\}
    \end{equation*}
    coincides with $\mathrm{U}(d_l)$.
    If the eigenspace additionally possesses a multi-partite structure ($d_l=2^k$), then $\mathscr{H}_l$ may be viewed as a $k$-qubit quantum code \cite{KL05,P05} on which universal manipulation of quantum information is possible in terms of holonomic gates $U_{A_l}(\gamma)$ only.

    A convenient measure of how close the group $\mathrm{Hol}(A_l)$ comes to span the entire unitary group is given in terms of the local curvature $F_l$ (the non-Abelian field-strength tensor). 
    It describes changes in the eigenstates in $\mathscr{H}_{l}$ under variation of the parameters $\kappa_\mu$. 
    Its antisymmetric components ($F_{l,\mu\nu}=-F_{l,\nu\mu}$) are calculated from \cite{N13}
    \begin{equation}
        \label{eq:curv}
        F_{l,\mu\nu}=\partial_{\mu}A_{l,\nu}-\partial_{\nu}A_{l,\mu}+[A_{l,\mu},A_{l,\nu}].
    \end{equation}
    According to a statement from differential geometry, the number of (linear-independent) components $\{F_{l,\mu\nu}\}_{\mu\nu}$ gives a lower bound to the dimension of $\mathrm{Hol}(A_{l})$.
    Here, dimension refers to the degrees of freedom that completely specify an element in a matrix group.
    For example, a unitary in $\mathrm{U}(d_l)$ is completely determined by specifying $d_l^2$ real numbers.
    Hence, we write $\mathrm{dim}\,\mathrm{U}(d_l)=d_l^2$.
    This implies that, if there are $d_l^2$ linear-independent matrices $F_{l,\mu\nu}$, it is possible to realise any element of the unitary group in terms of Eq. \eqref{eq:Hol} \cite{AS53,F00}, i.e., $\mathrm{Hol}(A_l)=\mathrm{U}(d_l)$.

    A more accurate bound on the dimension of $\mathrm{Hol}(A_l)$ can be obtained by including higher-order covariant derivatives
    \begin{equation}
    \label{eq:algebra}
        \nabla_{l,\sigma}F_{l,\mu\nu},~ \nabla_{l,\delta}\nabla_{l,\sigma}F_{l,\mu\nu},~ \nabla_{l,\epsilon}\nabla_{l,\delta}\nabla_{l,\sigma}F_{l,\mu\nu},~ \dots.
    \end{equation}
    Here, the covariant derivative operator
    \begin{equation*}
        \nabla_{l,\sigma}=\partial_\sigma+[A_{l,\sigma},\,\cdot\,]
    \end{equation*}
    generally is different for each eigenspace, thus depending on the index $l$.
    The number of linearly independent matrices in Eqs. \eqref{eq:curv} and \eqref{eq:algebra} equals the dimension of $\mathrm{Hol}(A_l)$ \cite{L02,L05}.
    
    Clearly, if the components $A_{l,\mu}$ are Abelian, then $\nabla_{l,\sigma}=\partial_\sigma$, and the span of the matrices $\{F_{l,\mu\nu},\partial_{\sigma}F_{l,\mu\nu},\dots\}_{\mu\nu\sigma\dots}$ is one-dimensional. 
    It follows that $\mathrm{Hol}(A_l)$ is an Abelian subgroup of $\mathrm{U}(d_l)$.
    Note that even though the above statements do not provide an explicit recipe for designing specific transformations, their existential nature makes them suitable for estimating the general potency of a quantum system to generate holonomies. 
    The dimension of the holonomy group acts as a natural measure of this potency.  

    \subsection{Four-mode fully-connected graph}
    \label{ssec:square}
    In order to illustrate the, rather abstract techniques introduced in the previous section, we give an example of a four-mode fully-connected graph, shown in Fig. \ref{fig:Mods} (c). 
    Fully-connected graphs constitute the most general type of graphs. 
    Hence, it is not expected that their Hamiltonians possess degenerate eigenvalues when arbitrary configurations $\bm{\kappa}=(\kappa_\mu)_{\mu=1}^M$ are considered. 
    Nevertheless, one can always construct specific configurations that lead to degenerate subspaces. 
    This is done as follows. 
    
    Let $\hat{H}_0$ be a time-independent Hamiltonian with some fixed degeneracy structure. 
    Consider the isospectral Hamiltonian
    \begin{equation}
    \label{eq:4FCG}
    \hat{H}(\bm{\kappa})=\hat{\mathcal{V}}(\bm{\kappa})\hat{H}_0 \hat{\mathcal{V}}^\dagger(\bm{\kappa}),
    \end{equation}
    parameterised over points $\bm{\kappa}=(\bm{\theta},\bm{\varphi})$ in $\mathscr{M}$. 
    For the four-mode system, let $\hat{H}_0=\hat{n}_1+\hat{n}_2-\hat{n}_4$ (with $\hat{n}_k=\hat{a}_k^\dagger\hat{a}_k$) and
    \begin{equation}
        \label{eq:unitary}
        \hat{\mathcal{V}}(\bm{\theta},\bm{\varphi})=\hat{V}_{12}(\theta_1,\varphi_1)\hat{V}_{23}(\theta_2,\varphi_2)\hat{V}_{34}(\theta_3,\varphi_3)
    \end{equation}
    is our unitary of choice.
    Here, $\hat{V}_{kk+1}(\theta_k,\varphi_k)$ creates a mixing between the modes $k$ and $k+1$. 
    More specifically, we define 
    \begin{equation}
        \label{eq:two-mode}
        \begin{split}
        \hat{V}_{kk+1}\hat{a}_k^\dagger \hat{V}_{kk+1}^\dagger&=\cos\theta_k\me^{\mi\varphi_k} \hat{a}_k^\dagger+\sin\theta_k\hat{a}_{k+1}^\dagger,\\
        \hat{V}_{kk+1}\hat{a}_{k+1}^\dagger\hat{V}_{kk+1}^\dagger&=\cos\theta_k\me^{-\mi\varphi_k}\hat{a}_{k+1}^\dagger-\sin\theta_k\hat{a}_{k}^\dagger,\\
    \end{split}
    \end{equation}
    which describes a general $\mathrm{SU}(2)$ transformation. 
    The transformation \eqref{eq:unitary} is not the most general unitary, but is chosen such that the Hamiltonian \eqref{eq:4FCG} is still bilinear in the creation and annihilation operators. 
    Thus, it can be represented by the graph in Fig. \ref{fig:Mods} (c). 
    
    If a single particle is subjected to the system, the Hamiltonian has a $4\times 4$ matrix representation $\hat{H}|_{\mathscr{F}_1}$. 
    Here, $\mathscr{F}_1$ denotes the first Fock layer, which contains the single-particle states $\ket{1_k}=\hat{a}^\dagger_{k}\ket{\bm{0}}$.
    In this Fock layer the system has only a single dark state, 
    \begin{equation}
    \label{eq:D1}
        \begin{split}
            \ket{D}&=\me^{\mi\varphi_3}\cos\theta_3(\sin\theta_1\sin\theta_2\ket{1_1}+\me^{-\mi\varphi_1}\cos\theta_1\sin\theta_2\ket{1_2}\\
            &\quad+\me^{-\mi\varphi_2}\cos\theta_2\ket{1_3})-\sin\theta_3\ket{1_4}.
        \end{split}
    \end{equation}
    A straightforward calculation of the corresponding connection [cf. Eq. \eqref{eq:con}] reveals (we omit the index $l=0$ for notational ease) 
    \begin{equation*}
        \begin{split}
        A_{\varphi_1}&=-\mi\cos^2\theta_1\sin^2\theta_2\cos^2\theta_3,\\
        A_{\varphi_2}&=-\mi\cos^2\theta_2\cos^2\theta_3,\\ 
        A_{\varphi_3}&=\mi\cos^2\theta_3,\\ 
        \end{split}
    \end{equation*}
    and $A_{\theta_1}=A_{\theta_2}=A_{\theta_3}=0$.
    The curvature is readily calculated from to Eq. \eqref{eq:curv}. Its nonvanishing components are
    \begin{equation*}
        \begin{split}
        F_{\varphi_1\theta_1}&=-2\mi \sin\theta_1\cos\theta_1\sin^2\theta_2\cos^2\theta_3,\\
        F_{\varphi_1\theta_2}&=2\mi \cos^2\theta_1\sin\theta_2 \cos\theta_2\cos^2\theta_3,\\
        F_{\varphi_1\theta_3}&=-2\mi \cos^2\theta_1\sin^2\theta_2\sin\theta_3 \cos\theta_3,\\
        F_{\varphi_2\theta_2}&=-2\mi \sin\theta_2\cos\theta_2 \cos^2\theta_3,\\
        F_{\varphi_2\theta_3}&=-2\mi \cos^2\theta_2 \sin\theta_3 \cos\theta_3,\\
        F_{\varphi_3\theta_3}&=2\mi \sin\theta_3 \cos\theta_3.\\
        \end{split}
    \end{equation*}
    It follows that Abelian holonomies (i.e., Berry phases) can be designed by adiabatically traversing loops in $\mathscr{M}$, i.e., $\mathrm{Hol}(A)=\mathrm{U}(1)$.

    Next, consider the second Fock layer $\mathscr{F}_{2}$ spanned by the two-particle states
    \begin{equation*}
        \begin{split}
        &\ket{2_1},~\ket{1_11_2},~\ket{1_11_3},~\ket{1_11_4},~\ket{2_2},\\
        &\ket{1_21_3},~\ket{1_21_4},~\ket{2_3},~\ket{1_31_4},~\ket{2_4}.\\
        \end{split}
    \end{equation*}
    The matrix $\hat{H}|_{\mathscr{F}_2}$ supports a three-fold degenerate dark subspace with states $\ket{D_k}$, for $k=1,2,3$ (explicit form in Appendix \ref{app:DS}).
    The connection on this subspace is 
    \begin{equation*}
        \begin{split}
        A_{\varphi_1}|_{\theta_1=\theta_3=\frac{\pi}{4}}^{\theta_2=0}&=\frac{\mi}{2}\begin{bmatrix}
        0&0&0\\
        0&1&\me^{\mi(\varphi_1-\varphi_2)}\\
        0&\me^{-\mi(\varphi_1-\varphi_2)}&-1\\
        \end{bmatrix},\\
        A_{\varphi_2}&=\mi\cos^2\theta_2\cos^2\theta_3
        \begin{bmatrix}
        -2&0&0\\
        0&-1&0\\
        0&0&1\\
        \end{bmatrix},\\ 
        A_{\varphi_3}&=\mi\cos^2\theta_3
        \begin{bmatrix}
        2&0&0\\
        0&-1&0\\
        0&0&1\\
        \end{bmatrix},\\
        A_{\theta_1}&=\cos\theta_2\begin{bmatrix}
        0&0&0\\
        0&0&-\me^{\mi(\varphi_2-\varphi_1)}\\
        0&\me^{\mi(\varphi_2-\varphi_1)}&0\\
        \end{bmatrix},\\
        \end{split}
    \end{equation*}
    and $A_{\theta_2}=A_{\theta_3}=0$.
    Calculating the curvature \eqref{eq:curv} and its first order covariant derivative gives rise to (only linearly independent components are shown)
    \begin{equation}
    \label{eq:C2}
        \begin{split}
        F_{\varphi_1\theta_1}|_{\bm{\kappa}_0}&=\begin{bmatrix}
        -\mi&0&0\\
        0&\frac{\mi}{2}&-\frac{\mi}{2\sqrt{2}}\\
        0&-\frac{\mi}{2\sqrt{2}}&-\frac{\mi}{2}\\
        \end{bmatrix},\\
        F_{\varphi_1\theta_2}|_{\bm{\kappa}_0}&=\begin{bmatrix}
        \mi&0&0\\
        0&\frac{\mi}{2}&\frac{\mi}{\sqrt{2}}\\
        0&\frac{\mi}{\sqrt{2}}&-\frac{\mi}{2}\\
        \end{bmatrix},\\
        F_{\varphi_2\theta_1}|_{\bm{\kappa}_0}&=\begin{bmatrix}
        0&0&0\\
        0&0&\mi\\
        0&\mi&0\\
        \end{bmatrix},\\
        \nabla_{\varphi_1}F_{\theta_1\theta_2}|_{\bm{\kappa}_0}&=\begin{bmatrix}
        0&0&0\\
        0&\mi&-\frac{\mi}{2\sqrt{2}}\\
        0&-\frac{\mi}{2\sqrt{2}}&-\mi\\
        \end{bmatrix},\\ 
        \nabla_{\theta_1}F_{\varphi_2\theta_1}|_{\bm{\kappa}_0}&=\begin{bmatrix}
        0&0&0\\
        0&-\mi&0\\
        0&0&\mi\\
        \end{bmatrix},\\
        \end{split}
    \end{equation}
    evaluated at the point $\bm{\kappa}_0$, with $\varphi_k=0$ and $\theta_k=\pi/4$.
    The matrices in Eq. \eqref{eq:C2} are the (infinitesimal) generators \cite{CL84} of a five-dimensional Lie group.
    This constitutes a lower bound to the dimension of $\mathrm{Hol}(A)$.
    Nevertheless, the analysis illustrates that the two-particle case enables the generation of more intriguing holonomies than the single-particle case.
    More precisely, the two-particle dark states led to a non-Abelian holonomy group $\mathrm{Hol}(A)$, which is a proper subgroup of $\mathrm{U}(3)$. 

    The key observation is that, increasing the number of particles significantly improved the computational capacity (from Abelian to non-Abelian holonomies) to generate unitaries on the dark subspace. 
    Intuitively, it is clear that, the dimension of $\mathrm{Hol}(A)$ cannot increase continually when the particle number becomes larger, as this would result in arbitrarily high computational power, while having only limited physical resources in $\mathscr{M}$.
    This leads us to an interesting question: 
    
    \textit{How far can one increase the dimension of the holonomy group by subjecting a larger number of particles to a system?} 
    
    This question will be addressed in the following section by means of a \textit{particle-number threshold}, which constitutes a formal answer to the issue.

    \section{Particle-number threshold}
    \label{sec:PNT}
    The previously presented benchmark system revealed the dependence of a system's holonomy group on the particle number $N$. 
    Firstly, this is due to the fact that the spectral properties (in particular  degeneracy) of a quantum system vary when the corresponding Hamiltonian $\hat{H}$ is limited to act on different Fock layers 
    \begin{equation*}
        \begin{split}
            \mathscr{F}_N&=\Big\{\ket{n_1,n_2,\dots}\,\Big|\,\sum\nolimits_{k} n_k=N\Big\}.\\
        \end{split}
    \end{equation*}
    Secondly, we noticed that even if the degeneracy increases, this does not necessarily mean that it is possible to generate a more useful (i.e., higher dimensional) subgroup of unitaries.
    Therefore, it is a natural question to ask, what is the particle number $N$ at which one of the holonomy groups $\{\mathrm{Hol}(A_l)\}_l$ reaches its maximal dimension and is therefore the most suitable for designing a versatile set of unitaries?  
    We refer to the number of particles necessary for this endeavour as the particle-number threshold (PNT) $\Nt$.   
    \begin{definition}
    Let $\hat{H}$ be the Hamiltonian of a quantum system in second quantisation that evolves adiabatically in time. 
    The particle-number threshold $\Nt$ denotes the minimum number of particles necessary
    to initialise any state in the eigenspace $\mathscr{H}_{l^\prime}$ whose holonomy group $\mathrm{Hol}(A_{l^\prime})$ has the highest dimension, i.e., 
    \begin{equation*}
        \mathrm{dim}\,\mathrm{Hol}(A_{l^\prime})\geq \mathrm{dim}\,\mathrm{Hol}(A_{l})
    \end{equation*}
    for all $l$ labeling the other eigenspaces $\mathscr{H}_l$ of $\hat{H}$.
    \end{definition}
    In contrast to previous examples, where the focus was on the dark subspace, finding the PNT $\Nt$ of a system demands an analysis of the holonomy groups of each eigenspace $\mathscr{H}_l$, in order to compute $\mathrm{dim}\,\mathrm{Hol}(A_l)$ for all $l$. 

    Intuitively speaking, the highest-dimensional holonomy group $\mathrm{Hol}(A_{l^\prime})$ is the one most useful for manipulating quantum states by geometric means. 
    In order to harness these transformations, we must be able to prepare quantum information in the corresponding eigenspace $\mathscr{H}_{l^\prime}$. 
    As states $\ket{\psi}$ in $\mathscr{H}_{l^\prime}(\bm{\kappa}_0)$ contain at most $\Nt$ particles, i.e., $\braket{\psi|\hat{n}|\psi}\leq \Nt$, one has to be able to prepare this particle number to fully harness the holonomy group's potential.   
    In the language of holonomic quantum computation \cite{ZR99,PZ99} the PNT of a quantum system $\hat{H}$ gives the number of particles to be prepared, in order to come as close as possible to the desirable notion of universality.

    \subsection{Properties of PNTs}
    \label{ssec:prop}
    The PNT $\Nt$ of a (bosonic) quantum system $\hat{H}$ is, in general, hard to calculate, as it demands for a calculation of the connection $A_l$ for each eigenspace (there could be infinitely many).
    Nevertheless, some general remarks can still be made.
    Consider a quantum system that consists of a collection of noninteracting subsystems, i.e., $\hat{H}=\bigotimes_{a}\hat{H}_a$. 
    Suppose that the PNT $\Nt^{(a)}$ for each subsystem $\hat{H}_a$ is known and that,
    $\mathrm{Hol}\big(A_{l^\prime}^{(a)}\big)$ denotes its holonomy group with maximal dimension.
    The composite system $\hat{H}$ then has PNT $\Nt=\sum_a \Nt^{(a)}$. 
    This becomes evident when noting that the highest-dimensional holonomy group 
    \begin{equation}
        \label{eq:HT}
        \mathrm{Hol}(A_{l^\prime})=\bigotimes_a \mathrm{Hol}\big(A_{l^\prime}^{(a)}\big)
    \end{equation}
    is just the tensor product of the holonomy groups $\mathrm{Hol}\big(A_{l^\prime}^{(a)}\big)$ of each individual subsystem. 
    The holonomy group \eqref{eq:HT} of the composite system acts on the subspace with energy $\prod_a \varepsilon_{l^\prime}^{(a)}$, where $\varepsilon_{l^\prime}^{(a)}$ denotes the eigenenergy of the subspace on which the group $\mathrm{Hol}\big(A_{l^\prime}^{(a)}\big)$ acts. 

    Next, consider a Hamiltonian with isospectral parametrisation, that is, 
    \begin{equation}
        \label{eq:Iso}
        \hat{H}(\bm{\kappa})=\hat{\mathcal{V}}(\bm{\kappa})\hat{H}_0\hat{\mathcal{V}}^\dagger(\bm{\kappa}),
    \end{equation}
    with $\hat{H}_0$ being a Hamiltonian with fixed degeneracy structure $\{d_l\}_l$ and eigenstates $\{\ket{\psi_{l,a}}\}_{l,a}$. 
    Suppose there is a sufficiently large parameter space $\mathscr{M}$ such that $\hat{\mathcal{V}}(\bm{\kappa})$ is the most general unitary operator.
    Adiabatic evolution in the $l$th eigenspace is then governed by the most general connection
    \begin{equation}
    \label{eq:con2}
        (A_{l,\mu})_{ab}=\bra{\psi_{l,b}}\hat{\mathcal{V}}^\dagger\partial_\mu \hat{\mathcal{V}}\ket{\psi_{l,a}}.
    \end{equation}
    In the above, we made use of the fact that $\hat{\mathcal{V}}(\bm{\kappa})\ket{\psi_{l,a}}$ are the eigenstates of \eqref{eq:Iso}.
    By construction, one has $\mathrm{Hol}(A_l)=\mathrm{U}(d_l)$. 
    For such a general parametrisation, it is, indeed, the eigenspace with the largest degeneracy $d_{l^\prime}\geq d_l$ that is the one most desirable for the generation of non-Abelian holonomies. 
    Hence, $\Nt$ is the number of particles necessary to populate any state in the most degenerate eigenspace $\mathscr{H}_{l^\prime}$.  

    \subsection{PNT of the Kerr-medium Hamiltonian}
    \label{ssec:Kerr}
    What happens when $\hat{\mathcal{V}}$ is not an arbitrary unitary, but is limited to some smaller set of physically accessible operations? 
    For concreteness, consider the two-mode Hamiltonian associated with a nonlinear Kerr medium
    \begin{equation*}
        \hat{H}_0=\hat{n}_1(\hat{n}_1-\hat{1})+\hat{n}_2(\hat{n}_2-\hat{1}).
    \end{equation*}
    Here, the unitary $\hat{\mathcal{V}}(\alpha,\beta,\xi,\zeta)$ is a product of single and two-mode displacement
    \begin{equation}
        \label{eq:Dis}
        \begin{split}
        \hat{D}_k(\alpha)&=\mathrm{exp}\big(\alpha\hat{a}_{k}^\dagger-\alpha^*\hat{a}_{k}\big),\\
        \hat{K}(\beta)&=\mathrm{exp}\big(\beta\hat{a}_{1}^\dagger \hat{a}_{2}-\beta^*\hat{a}_{1} \hat{a}_{2}^\dagger\big),
        \end{split}
    \end{equation}
    as well as single and two-mode squeezing 
    \begin{equation}
        \label{eq:Squ}
        \begin{split}
        \hat{S}_k(\xi)&=\mathrm{exp}\big(\xi(\hat{a}_{k}^\dagger)^2-\xi^*\hat{a}_{k}^2\big),\\
        \hat{M}(\zeta)&=\mathrm{exp}\big(\zeta\hat{a}_{1}^\dagger \hat{a}_{2}^\dagger-\zeta^*\hat{a}_{1} \hat{a}_{2}\big),
        \end{split}
    \end{equation}
    respectively \cite{V06}. 
    By driving coherent displacement $(\alpha,\beta)$ and squeezing parameters $(\xi,\zeta)$ through a closed loop in $\mathscr{M}=\mathbb{C}^4$, holonomies on the eigenspaces of $\hat{H}$ are obtained.
    In Ref. \cite{PC00} it was shown that this enables arbitrary $\mathrm{U}(4)$ transformations over the zero-eigenvalue eigenspace $\mathscr{H}_0$.
    This was done by explicitly constructing loops that implement the square root of a swap gate together with a holonomic single-qubit rotation. In Ref. \cite{L02} the author comes to the same conclusion but via an analysis of the curvature and its covariant derivatives.
    Note that the subspace $\mathscr{H}_0$ [at the base point $(\alpha,\beta,\xi,\zeta)=\bm{0}$] is spanned by the number states $\ket{0_10_2}$, $\ket{1_10_2}$, $\ket{0_11_2}$, and $\ket{1_11_2}$; that is, two photons are necessary to initialise any state in the subspace.
    
    In order to determine the PNT of the Kerr-medium Hamiltonian, one has to check whether the higher-energy eigenspaces offer any computational advantage; that is, do we find a holonomy group $\mathrm{Hol}(A_l)>\text{U}(4)$?
    Let us make the first step of this analysis explicit.
    Given a maximum of three photons, each of the states $\ket{0_1,2_2}$, $\ket{1_1,2_2}$, $\ket{2_1,1_2}$, and $\ket{2_1,0_2}$, spanning the eigenspace $\mathscr{H}_1$ with energy $\varepsilon_1=2$, can be initialised. 
    Starting from the connection \eqref{eq:con2} for $\hat{\mathcal{V}}=\hat{K}(\beta)\hat{M}(\zeta)\hat{D}_k(\alpha)\hat{S}_j(\xi)$, the curvature \eqref{eq:curv} and its covariant derivatives can be calculated (see Appendix \ref{app:curv}). 
    We find $16$ linearly-independent matrices, thus the connection on this subspace is irreducible, i.e., $\mathrm{Hol}(A_1)=\mathrm{U}(4)$. 
    We conclude that the holonomy group $\mathrm{Hol}(A_1)$ does not offer any advantage over $\mathrm{Hol}(A_0)$, but demands for the preparation of an additional photon.
    In order to evaluate the PNT of the system, the analysis has to be continued for higher energy eigenspaces $\mathscr{H}_l$ (with $l\geq 2$), which demands for the preparation of a higher photon number $N$.

    An extended study (up to $N=50$) of the curvature $F_l$ shows that, even though further increasing the particle number ($N>3$) populates subspaces with increased degeneracy (up to $d_l=10$ for some eigenspaces), their holonomy groups do not offer a computational advantage. 
    By that we mean 
    \begin{equation*}
        \mathrm{dim}\,\mathrm{Hol}(A_l)\leq \mathrm{dim}\,\mathrm{Hol}(A_0)
    \end{equation*}
    verified for all eigenspaces $\mathscr{H}_l$ with index $l\leq 352$ (cf. Tab. \ref{tab:Kerr}).
    We did so by explicitly calculating the components $F_{l,\mu\nu}$ of the curvature and their covariant derivatives up to order $3$ (these are too large to display here).
    The computed dimension of the groups $\{\mathrm{Hol}(A_l)\}_l$ did not increase further after the first order derivatives, thus giving us good confidence that the dimension was determined accurately. 

    There is an intuitive explanation for the fact that eigenspaces involving higher particle numbers ($N>6$) lead to less useful holonomy groups.
    The Gaussian operations \eqref{eq:Dis} and \eqref{eq:Squ} contribute to the evolution only via the connection $A_l$.
    The derivative $\partial_\mu$ in Eq. \eqref{eq:con2} that acts on the operators \eqref{eq:Dis} and \eqref{eq:Squ}, leads to creation and annihilation operators of (at most) quadratic order.
    Hence, Fock states with larger differences in their photon numbers cannot be transformed into each other by a quantum holonomy, even when they lie in the same subspace.
    
    In summary, the subspace $\mathscr{H}_0$ (containing at most two-particle states) should be preferred when the system is utilised in a holonomic quantum computation.
    Therefore, the PNT of the two-mode Kerr Hamiltonian is $\Nt=2$. 
    Moreover, it was shown that restricting the parametrisation of the Hamiltonian \eqref{eq:Iso} to unitaries $\hat{\mathcal{V}}$ that can be implemented by Gaussian operations \eqref{eq:Dis} and \eqref{eq:Squ}, led to most of the system's eigenspaces having reducible connections $A_l$, i.e., $\mathrm{Hol}(A_l)\subset\mathrm{U}(d_l)$. 
    Hence, degeneracy became a quantity of secondary interest.
    In Tab. \ref{tab:Kerr} the spectral properties of the two-mode Kerr Hamiltonian $\hat{H}$ are listed together with their capacity to generate holonomies on the eigenspaces $\mathscr{H}_l$ (for $l=0,\dots,352$).
    Note that subspaces with degeneracy $d_l\leq 4$ are not listed in in Tab. \ref{tab:Kerr}, as it is already clear that their holonomy groups cannot exceed the dimension of $\mathrm{Hol}(A_0)=\mathrm{U}(4)$.

    \begin{table}[h]
    \centering
    \begin{tabular}{c c c c c c}
    \hline\hline
    $\quad l\quad$&$\quad\varepsilon_l\quad$&$\quad d_l\quad$&$\quad\leq N\quad$&$\mathrm{dim}\{F_{l,\mu\nu}\}_{\mu\nu}$&$\mathrm{dim}\,\mathrm{Hol}(A_l)$ \\\hline
     0 & 0 & 4 & 2 & 14 & 16\\
     1 & 2 & 4 & 3 & 14 & 16\\
     5 & 12 & 5 & 6 & 9 & 9\\
     16 & 42 & 6 & 10 & 9 & 9\\
     26 & 72 & 6 & 13 & 12 & 12\\
     37 & 110 & 6 & 15 & 9 & 9\\
     45 & 132 & 6 & 17 & 9 & 9\\
     54 & 162 & 6 & 19 & 6 & 6\\
     60 & 182 & 6 & 20 & 9 & 9\\
     70 & 212 & 6 & 21 & 3 & 3\\
     78 & 240 & 6 & 21 & 9 & 9\\
     87 & 272 & 6 & 24 & 9 & 9\\
     99 & 312 & 5 & 26 & 3 & 3\\
     108 & 342 & 6 & 27 & 9 & 9\\
     113 & 362 & 6 & 27 & 3 & 3\\
     130 & 420 & 5 & 30 & 9 & 9\\
     131 & 422 & 6 & 30 & 3 & 3\\
     141 & 462 & 8 & 31 & 9 & 9\\
     157 & 512 & 6 & 33 & 6 & 6\\
     168 & 552 & 10 & 34 & 9 & 9\\
     199 & 662 & 6 & 36 & 3 & 3\\
     208 & 702 & 6 & 38 & 9 & 9\\
     215 & 722 & 6 & 39 & 6 & 6\\
     222 & 756 & 6 & 38 & 9 & 9\\
     225 & 762 & 6 & 40 & 3 & 3\\
     238 & 812 & 10 & 41 & 9 & 9\\
     266 & 912 & 6 & 42 & 3 & 3\\
     274 & 942 & 8 & 44 & 3 & 3\\
     285 & 992 & 6 & 45 & 9 & 9\\
     306 & 1062 & 8 & 47 & 3 & 3\\
     320 & 1112 & 6 & 48 & 3 & 3\\
     323 & 1122 & 6 & 48 & 9 & 9\\
     346 & 1202 & 8 & 50 & 3 & 3\\
     349 & 1212 & 6 & 49 & 3 & 3\\
     352 & 1232 & 8 & 50 & 3 & 3\\
         \hline\hline
    \end{tabular}
    \caption{\label{tab:Kerr} Holonomy groups of the two-mode nonlinear Kerr medium         parameterised by the Gaussian operations \eqref{eq:Dis} and \eqref{eq:Squ}. The     table contains the degeneracy $d_l$ of the $l$th eigenspace (with energy            $\varepsilon_l$).
        $N$ denotes the number of particles necessary to fully occupy the corresponding eigenspace.
        The number of linear-independent curvature components $F_{l,\mu\nu}$ as well as the dimension of the holonomy group $\mathrm{Hol}(A_l)$. 
        Covariant derivatives were calculated up the order of $3$.}
    \end{table}

    \subsection{PNTs of coupled harmonic oscillators}
    \label{ssec:LO}
    While the exact calculation of a PNT can be a daunting task, given a collection of coupled harmonic oscillators, certain specialisations arise that can simplify calculations drastically.
    In Fig. \ref{fig:Mods} such systems were represented as graphs.
    The calculation of PNTs for such systems would be relevant, for instance, to the geometric manipulation of multi-photon states in linear optics \cite{PT20} as well as linear optical quantum computation by holonomic means \cite{PS22}. 

    Population transfer between different Fock layers $\mathscr{F}_{N}$ does not occur in these systems, as the total number of particles stays conserved throughout an evolution. 
    From a mathematical viewpoint, this implies that the system's Hamiltonian reveals a block-matrix structure, i.e., 
    \begin{equation*}
        \hat{H}=\bigoplus_{N\in\mathbb{N}}\hat{H}|_{\mathscr{F}_N}.
    \end{equation*}
    In addition, there always exists a spectral decomposition $\hat{H}=\sum_l \varepsilon_l\hat{\Pi}_l$, with $\hat{\Pi}_l$ denoting the projector onto the eigenspace $\mathscr{H}_l$. 
    It follows that the eigenspaces themselves admit a similar decomposition, that is,
    \begin{equation}
        \label{eq:dec2}
        \hat{\Pi}_{l}=\bigoplus_{N(l)}\hat{\Pi}_{l}|_{\mathscr{F}_{N(l)}},
    \end{equation}
    where summation is carried out over those particle numbers $N(l)$ at which the corresponding energy $\varepsilon_l$ occurs. 

    As an example, the Hamiltonian \eqref{eq:H1} of the $\Lambda$-scheme [Fig. \ref{fig:Mods}. (a)] does not possess single-particle eigenstates with energy $2\sqrt{|\kappa_+|^2+|\kappa_-|^2}$.
    In other words, the eigenvalue does not lie in the spectrum of $\hat{H}|_{\mathscr{F}_1}$, but it is an eigenvalue of the matrix $\hat{H}|_{\mathscr{F}_N}$ for $N\geq2$. 
    In this case, the sum in Eq. \eqref{eq:dec2} corresponds to an infinite series starting with $N(l)=2,3,\dots$.

    If additionally the evolution is assumed to be adiabatic, population transfer occurs within each eigenspace separately.
    Hence, the decomposition \eqref{eq:dec2} is inherited to the time-evolution operator (quantum holonomy)
    \begin{equation}
        \label{eq:dec3}
        U_{A_l}(\gamma)=\bigoplus_{N(l)} U_{A_l}(\gamma)|_{\mathscr{F}_{N(l)}}.
    \end{equation}
    Remarkably, the connection will always be reducible for such a system, because it is not possible to generate transformations between different Fock layers.
    The best one can hope for is to find is a highly-degenerate $N$-particle block in the eigenspace $\mathscr{H}_l$ such that the holonomy $U_{A_l}(\gamma)|_{\mathscr{F}_{N}}$ realises any unitary transformation on the subspace $\mathscr{H}_l|_{\mathscr{F}_{N}}$.
    This is nothing but a geometric incarnation of the well-known fact that networks of coupled oscillators (by themselves) do not allow for universal quantum computation \cite{KM07}, but must be supported by additional resources, such as measurement-induced nonlinearities \cite{KLM,SN03}.

    Note that even though the quantum holonomy \eqref{eq:dec3} can have an infinite-dimensional matrix representation, it might still be commuting, that is $U_{A_l}(\gamma)U_{A_l}(\gamma^\prime)=U_{A_l}(\gamma^\prime)U_{A_l}(\gamma)$ for any two loops $\gamma$ and $\gamma^\prime$ in $\mathscr{M}$. 
    For the purpose of illustration, consider the Hamiltonian \eqref{eq:H1} of the $\Lambda$-scheme [cf. Fig. \ref{fig:Mods} (a)] which gives rise to an infinite-dimensional dark subspace.
    For a single photon, the matrix $\hat{H}|_{\mathscr{F}_1}$ has only one dark state.
    Given two or three photons in the setup, $\hat{H}|_{\mathscr{F}_2}$ and $\hat{H}|_{\mathscr{F}_3}$ both have two dark states. 
    Subjecting four photons to the system leads to a Hamiltonian matrix
    $\hat{H}|_{\mathscr{F}_4}$ having three dark states. 
    Even though degeneracy further increases, the quantum holonomy
    \begin{equation*}
        U_{A_0}=\begin{bmatrix}
        U_{A_0}|_{\mathscr{F}_{1}}&&\\
        &U_{A_0}|_{\mathscr{F}_{2}}&\\
        &&\ddots\\
        \end{bmatrix}
    \end{equation*}
    will remain Abelian, because the $N$-particle block
    \begin{equation*}
        U_{A_0}(\gamma)|_{\mathscr{F}_{N}}=\mathrm{diag}\Big(\me^{\mi N\phi(\gamma)},\dots,\me^{-\mi N\phi(\gamma)}\Big),
    \end{equation*}
    is itself a diagonal matrix [cf. Eq. \eqref{eq:2LH} for $N=2$].
    Here, $\phi(\gamma)$ is the geometric phase defined in Eq. \eqref{eq:GP}.
    The above analysis illustrates, that increasing the particle number in the photonic $\Lambda$-scheme does not increase the holonomy group's dimension; that is, it stays Abelian. 
    Similar arguments hold for the other eigenspaces of the system, and thus, a single photon is sufficient to generate any phase in $\mathrm{U}(1)$.
    Hence, the PNT is $\Nt=1$.

    \subsubsection{Three-mode fully-connected graph}
    Consider a setup containing three oscillator modes $\hat{a}_\pm$ and $\hat{a}_0$.
    Coupling between the modes is described by the parameters $\kappa_\pm$ and $\kappa_0$, respectively.
    The system corresponds to the three-mode fully-connected graph shown in Fig. \ref{fig:Mods} (d). 
    
    For simplicity, its Hamiltonian is considered to be in the configuration
    \begin{equation}
    \label{eq:orbit2}
    \hat{H}(\bm{\theta},\bm{\varphi})=\hat{\mathcal{V}}(\bm{\theta},\bm{\varphi})\hat{H}_0 \hat{\mathcal{V}}^\dagger(\bm{\theta},\bm{\varphi}),
    \end{equation}
    with $\hat{H}_0=\hat{n}_+-\hat{n}_-$. 
    In the above,
    \begin{equation*}
        \hat{\mathcal{V}}(\bm{\theta},\bm{\varphi})=\hat{V}_{+0}(\theta_+,\varphi_+)\hat{V}_{0-}(\theta_-,\varphi_-),
    \end{equation*}
    with the operator $\hat{V}_{kk+1}$ defined in Eq. \eqref{eq:two-mode}.
    The $3\times 3$ matrix $\hat{H}|_{\mathscr{F}_1}$ possesses single-particle eigenstates
    \begin{equation*}
        \begin{split}
        \ket{B_+}&=\cos\theta_+ \me^{\mi\varphi_+}\ket{1_+}-\sin\theta_- \ket{1_0},\\
        \ket{D}&=\cos\theta_-\me^{\mi\varphi_-}\big(\cos\theta_+ \me^{-\mi\varphi_+}\ket{1_+}-\sin\theta_+ \ket{1_0}\big)\\
        &\quad-\sin\theta_- \ket{1_-},\\
        \ket{B_-}&=\cos\theta_-\me^{-\mi\varphi_-}\ket{1_-}\\
        &\quad-\sin\theta_-\big(\cos\theta_+ \me^{-\mi\varphi_+}\ket{1_0}-\sin\theta_+ \ket{1_+}\big),\\
        \end{split}
    \end{equation*}
    with eigenenergies $\varepsilon_\pm=\pm1$ and $\varepsilon_0=0$.
    The connection for each eigenvalue is readily calculated via Eq. \eqref{eq:con}, leading to the nonvanishing components
    \begin{equation*}
        \begin{split}
        A_{+,\varphi_{+}}&=\mi\cos^2\theta_{+},\hspace{2.05cm}
        A_{0,\varphi_{\pm}}=\mi \cos^2\theta_{\pm},\\
        A_{-,\varphi_{+}}&=-\mi\sin^2\theta_{+}\cos^2\theta_{+},\qquad A_{-,\varphi_{-}}=-\mi\cos^2\theta_{-}.\\
        \end{split}
    \end{equation*}

    When given more than just a single particle, the Hamiltonian \eqref{eq:orbit2} gives rise to degenerate subspaces; for example, considering two particles in the system, the $6\times6$ matrix $\hat{H}|_{\mathscr{F}_2}$ possesses two dark states.
    However, in the following it will be shown that the resulting holonomies are still Abelian for arbitrary particle numbers $N$.
    It is a well-known fact for coupled-mode systems, that knowing the single-particle evolution is equivalent to knowing the evolution of the modes $\hat{a}_k(t)$ in the Heisenberg picture \cite{V06}, viz., 
    \begin{equation*}
        \hat{a}_\pm^\dagger(T)=\me^{\mp\mi T}\me^{\oint A_\pm}\hat{a}_\pm^\dagger(0),\qquad\hat{a}_0^\dagger(T)=\me^{\oint A_0}\hat{a}_0^\dagger(0).\\
    \end{equation*}
    Subsequently, the evolution of any $N$-particle state can be given explicitly. 
    It follows that the state can attain only a Berry phase as well.
    The initial $N$-particle state 
    \begin{equation*}
        \ket{\Psi(0)}=\frac{1}{\sqrt{n_+!n_0!n_-!}}\big(\hat{a}_+^\dagger\big)^{n_+}\big(\hat{a}_0^\dagger\big)^{n_0}\big(\hat{a}_-^\dagger\big)^{n_-}\ket{\bm{0}}
    \end{equation*}
    ($N=n_++n_0+n_-$) adiabatically evolves into
    \begin{equation*}
        \ket{\Psi(T)}=\me^{n_+\oint A_+}\me^{n_0\oint A_0}\me^{n_-\oint A_-}\ket{n_+}\otimes\ket{n_0}\otimes\ket{n_-},
    \end{equation*}
    accumulating a (scalar) geometric phase. 
    
    We thus conclude that, independent of the provided particle number $N$, the holonomy group can be only Abelian. 
    Hence, the PNT of the system is $\Nt=1$, as moving beyond the single-particle case did not lead to more versatile groups of holonomies, just higher-dimensional representations of the group $\mathrm{U}(1)$.
    The above argument is the special case of a more general bosonic-operator framework,
    which we devised in Ref. \cite{PS22}.
    This formalism enables a photon-number-independent description of holonomies, and thus might be useful for the calculation of PNTs in coupled-mode systems.

    \subsection{PNTs of fermionic systems}
    \label{sec:Fermi}
    So far, all considered quantum systems were bosonic in nature. 
    Nevertheless, the definition of a PNT is applicable to any quantum system given in second quantisation (cf. Sec. \ref{sec:PNT}). 
    Fermionic modes are associated with creation and annihilation operators satisfying canonical anticommutation relations. 
    Because of this, the most prominent difference from the bosonic setups studied previously, is that fermions have to obey the Pauli principle, i.e., two fermions cannot occupy the same mode simultaneously. 
    This drastically reduces the number of possible states in a system, and in particular, the corresponding Hilbert space (Fock space) is finite-dimensional.
    Hence, the calculation of the PNT of a fermionic system becomes much more manageable in comparison to bosonic systems.

    PNTs can also be calculated for systems comprising both bosonic and fermionic modes. 
    As an elementary example, consider the Jaynes-Cummings Hamiltonian describing the interaction between an incident light field and a single atomic energy level at resonance. 
    Within the rotating wave approximation, the Hamiltonian reads \cite{V06}
    \begin{equation*}
        \hat{H}_{\text{JC}}=\omega_{\text{A}}\hat{\sigma}^+ \hat{\sigma}^- + \omega_{\text{c}}\hat{n}+\kappa\big(\hat{a}^\dagger\hat{\sigma}^- + \hat{a}\hat{\sigma}^+ \big),
    \end{equation*}
    with $\omega_{\text{A}}$ being the resonance frequency of the atom, $\omega_{\text{c}}$ being the frequency of the incident light field, and $\kappa$ describing the strength of the light-matter interaction. 
    The atomic ladder operators $\hat{\sigma}^-$ and $\hat{\sigma}^+=(\hat{\sigma}^-)^\dagger$ shift an electron from the ground to the excited state and vice versa.
    The system possesses a nondegenerate spectrum $\{\varepsilon_{n^\pm}\}_{n\in\mathbb{N}}$ with the corresponding eigenstates 
    \begin{equation*}
        \begin{split}
        \ket{n^+}&=\sin\theta\ket{g,n+1}+\cos\theta\ket{e,n},\\
        \ket{n^-}&=\cos\theta\ket{g,n+1}-\sin\theta\ket{e,n},\\
        \end{split}
    \end{equation*}
    where $\tan(2\theta)=2\kappa\sqrt{n+1}/(\omega_{\text{c}}-\omega_{\text{A}})$ and $n$ is the photon number.
    This form of the eigenstates highlights that the underlying parameter space does not possess any curvature, i.e., $F_{n^{\pm},\theta\theta}=0$ for all photon numbers $n\in\mathbb{N}$.
    Hence, the system is not suitable for the generation of quantum holonomies, and this is reflected in the PNT, i.e., $\Nt=0$.

    \section{Discussion}
    \label{sec:FIN}
    In this article we studied quantum holonomies in relation to the particle number in a system.
    It was shown that, increasing the number of particles can lead to a higher-dimensional holonomy group, thus improving the capabilities of the system to generate useful unitaries.
    We introduced the PNT of a quantum system, which denotes the minimal number of particles necessary to fully exploit the systems capacity for generating a versatile set of quantum holonomies.
    In addition to some general statements that could be made about PNTs, we illustrated the theory in terms of benchmark examples relevant to linear and nonlinear quantum optics.
    We saw that for systems of coupled oscillators only the $\Nt$-particle block of an eigenspace contributes to its holonomy group relevantly, because the particle number $\Nt$ subjected to the system does not change throughout the propagation. 
    This result appears to be relevant to linear optical quantum computation by adiabatic means. 
    We argued that the results presented are applicable to both bosonic and fermionic systems of interest.
    Our general investigation hints at the utility of the concept in assessing the capabilities of different quantum systems to perform holonomic quantum computations in terms of holonomies. 
    PNTs might also be relevant to the simulation of gauge groups in terms of adiabatic parameter variations. 

    Currently, there is a lack of analytical tools to compute PNTs.
    While this is a straight forward task for fermionic systems, it becomes a challenging issue for bosonic systems, where there is no bound on the particle number. 
    Estimations of the PNT up to some finite particle number, might be sufficient for most practical purposes, but general strategies for calculating PNTs can be relevant for a deeper understanding of many-particle physics in an adiabatic setting. 

    \acknowledgments
    Financial support from the Deutsche Forschungsgemeinschaft (DFG SCHE 612/6-1) is gratefully acknowledged.

    \appendix

    \section{Two-particle dark states of the four-mode fully-connected graph}
    \label{app:DS}
    In this appendix, we give the two-particle dark states of the Hamiltonian matrix $\hat{H}|_{\mathscr{F}_2}$ of the four-mode fully-connected graph shown in Fig. \ref{fig:Mods} (c). 
    The isospectral Hamiltonian of the system is given in Eq. \eqref{eq:4FCG}.
    The three dark states of the system read 
    \begin{widetext}
        \begin{equation*}
            \begin{split}
            \ket{D_1}&=\frac{1}{\sqrt{2}}\big(\me^{\mi\varphi_3}\sin\theta_1\sin\theta_2\cos\theta_3\hat{a}_1^\dagger+\me^{\mi(\varphi_3-\varphi_1)}\cos\theta_1\sin\theta_2\cos\theta_3\hat{a}_2^\dagger+\me^{\mi(\varphi_3-\varphi_2)}\cos\theta_2\cos\theta_3\hat{a}_3^\dagger-\sin\theta_3\hat{a}_4^\dagger\big)^2\ket{\bm{0}},\\
            \ket{D_2}&=\big(\sin\theta_1\sin\theta_2\sin\theta_3\hat{a}_1^\dagger+\me^{-\mi\varphi_1}\cos\theta_1\sin\theta_2\sin\theta_3\hat{a}_2^\dagger+\me^{-\mi\varphi_2}\cos\theta_2\sin\theta_3\hat{a}_3^\dagger+\me^{-\mi\varphi_3}\cos\theta_3\hat{a}_4^\dagger\big)\\
            &\quad\times\big(\me^{\mi\varphi_1}\cos\theta_1\hat{a}_1^\dagger-\sin\theta_1\hat{a}_2^\dagger\big)\ket{\bm{0}},\\
            \ket{D_3}&=\big(\sin\theta_1\sin\theta_2\sin\theta_3\hat{a}_1^\dagger+\me^{-\mi\varphi_1}\cos\theta_1\sin\theta_2\sin\theta_3\hat{a}_2^\dagger+\me^{-\mi\varphi_2}\cos\theta_2\sin\theta_3\hat{a}_3^\dagger+\me^{-\mi\varphi_3}\cos\theta_3\hat{a}_4^\dagger\big)\\
            &\quad\times\big(\me^{\mi\varphi_1}\sin\theta_1\cos\theta_2\hat{a}_1^\dagger+\me^{\mi(\varphi_2-\varphi_1)}\cos\theta_1\cos\theta_2\hat{a}_2^\dagger\big)\ket{\bm{0}},\\
            \end{split}
        \end{equation*}
    where the parameter angles $(\theta_k,\varphi_k)$ are defined in Eq. \eqref{eq:two-mode}.    
    \end{widetext}

    \section{Curvature on the subspace $\mathscr{H}_1$ of the Kerr-medium Hamiltonian}
    \label{app:curv}
    Given the isospectral family $\hat{H}=\hat{\mathcal{V}}\hat{H}_0 \hat{\mathcal{V}}^\dagger$ of the Kerr-medium Hamiltonian
    \begin{equation*}
        \hat{H}_0=\hat{n}_1(\hat{n}_1-\hat{1})+\hat{n}_2(\hat{n}_2-\hat{1}),
    \end{equation*}
    non-Abelian holonomies are generated over the eigenspace $\mathscr{H}_1$ (with energy $\varepsilon_1=2$) due to a mixing of the eigenstates
    \begin{equation*}
        \hat{\mathcal{V}}\ket{0_1,2_2},\quad\hat{\mathcal{V}}\ket{1_1,2_2},\quad\hat{\mathcal{V}}\ket{2_1,1_2},\quad\hat{\mathcal{V}}\ket{2_1,0_2},
    \end{equation*}
    where
    \begin{equation*}
        \hat{\mathcal{V}}(\bm{r},\bm{\theta})=\hat{K}\big(r_4\me^{\mi\theta_4}\big)\hat{M}\big(r_3\me^{\mi\theta_3}\big)\hat{D}_1\big(r_1\me^{\mi\theta_1}\big)\hat{S}_1\big(r_2\me^{\mi\theta_2}\big)
    \end{equation*}
    is a product of the single and two-mode operations in Eqs. \eqref{eq:Dis} and \eqref{eq:Squ}.
     
    Calculating the connection $A_1$ on this subspace via Eq. \eqref{eq:con2} reveals nonvanishing components
    \begin{widetext}
        \begin{equation*}
        A_{1,r_1}|_{(\bm{0},\zeta)}=\begin{bmatrix}
        0&-1&0&0\\
        1&0&0&0\\
        0&0&0&0\\
        0&0&0&0\\
        \end{bmatrix},\qquad
        A_{1,r_4}|_{(\bm{0},\zeta)}=\begin{bmatrix}
        0&0&0&0\\
        0&0&0&-2\me^{-\mi\theta_4}\\
        0&0&0&0\\
        0&2\me^{\mi\theta_4}&0&0\\
        \end{bmatrix},
        \end{equation*}
        \begin{equation*}
        A_{1,\theta_4}|_{(\bm{0},\zeta)}=\begin{bmatrix}
                    4\mi r_4s_1|s_2| & 0 & 0 & 0\\
                    0 & 2\mi r_4s_1|s_2| & 0 & 2\mi\zeta^*\cos(2r_4)\\
                    0 & 0 & -4\mi r_4s_1|s_2| & 0\\
                    0 & 2\mi\zeta\cos(2r_4) & 0 & -2\mi r_4s_1|s_2|\\
                \end{bmatrix},
    \end{equation*}
    where $s_1=\cos(|\zeta|)$, $s_2=\me^{\mi\theta_4}\sin(|\zeta|)$, and $\zeta=r_4\me^{\mi\theta_4}$.
    Above, we evaluated the connection at a point with $r_{1}=r_2=r_3=0$, and $\theta_{1}=\theta_2=\theta_3=0$.
    The corresponding curvature \eqref{eq:curv} [evaluated at $(\bm{0},\zeta)$] can be computed as
        \begin{equation*}
        F_{1,r_1r_2}|_{(\bm{0},\zeta)}=\begin{bmatrix}
        0&-2&0&0\\
        2&0&0&0\\
        0&0&0&0\\
        0&0&0&0\\
        \end{bmatrix},\quad 
        F_{1,r_1r_3}|_{(\bm{0},\zeta)}=\begin{bmatrix}
        0&0&0&0\\
        0&0&0&0\\
        0&0&0&-1\\
        0&0&1&0\\
        \end{bmatrix},\quad F_{1,r_2r_3}|_{(\bm{0},\zeta)}=\begin{bmatrix}
                    0 & 0 & 0 & 0\\
                    0 & 0 & 0 & 4\\
                    0 & 0 & 0 & 0\\
                    0 & -4 & 0 & 0\\
            \end{bmatrix},
        \end{equation*}
        \begin{equation*}
            F_{1,r_1r_4}|_{(\bm{0},\zeta)}=\begin{bmatrix}
                    0 & 0 & 0 & 2\me^{-\mi\theta_4}\\
                    0 & 0 & 0 & 0\\
                    0 & 0 & 0 & \me^{\mi\theta_4}\\
                    -2\me^{\mi\theta_4} & 0 & -\me^{-\mi\theta_4} & 0\\
                \end{bmatrix},\quad
            F_{1,r_1\theta_4}|_{(\bm{0},\zeta)}=\begin{bmatrix}
                    0 & 0 & 0 & -2\mi \zeta^*\cos(2r_4)\\
                    0 & 0 & 0 & 0\\
                    0 & 0 & 0 & \mi\zeta\cos(2r_4)\\
                    -2\mi \zeta\cos(2r_4) & 0 & \mi\zeta^*\cos(2r_4) & 0\\
                \end{bmatrix},
            \end{equation*}
            \begin{equation*}
                F_{1,r_4\theta_4}|_{(\bm{0},\zeta)}=\begin{bmatrix}
                    4\mi(r_4\cos(2r_4)+s_1|s_2|) & 0 & 0 & 0\\
                    0 & -2\mi(3r_4\cos(2r_4)-s_1|s_2|) & 0 & -4\mi|s_2|^2\me^{-\mi\theta_4}\\
                    0 & 0 & -2\mi(r_4\cos(2r_4)+s_1|s_2|) & 0\\
                    0 & -4\mi|s_2|^2\me^{\mi\theta_4}& 0 & 2\mi(3r_4\cos(2r_4)-s_1|s_2|)
                \end{bmatrix}.
        \end{equation*}
        Its first-order covariant derivatives are found to be 
        \begin{equation*}
                \nabla_{\theta_1}F_{1,r_1 r_2}|_{(\bm{0},\zeta)}=\begin{bmatrix}
                    0 & -2\mi & 0 & \\
                    -2\mi & 0 & 0 & 0\\
                    0 & 0 & 0 & 0\\
                    0 & 0 & 0 & 0\\
                \end{bmatrix},~\nabla_{r_4}F_{1,r_1 r_3}|_{(\bm{0},\zeta)}=\begin{bmatrix}
                    0 & 0 & 0 & 0\\
                    0 & 0 & -2\me^{-\mi\theta_4} & 0\\
                    0 & 2\me^{\mi\theta_4} & 0 & 0\\
                    0 & 0 & 0 & 0\\
                \end{bmatrix},~
                \nabla_{r_4}F_{1,r_1r_2}|_{(\bm{0},\zeta)}=\begin{bmatrix}
                    0 & 0 & 0 & -4\me^{-\mi\theta_4}\\
                    0 & 0 & 0 & 0\\
                    0 & 0 & 0 & 0\\
                    4\me^{\mi\theta_4} & 0 & 0 & 0\\
                \end{bmatrix},
            \end{equation*}
            \begin{equation*}
                \nabla_{r_1}F_{1,r_1\theta_1}|_{(\bm{0},\zeta)}=\begin{bmatrix}
                    2\mi & 0 & 0 & 0\\
                    0 & 6\mi & 0 & 0\\
                    0 & 0 & 4\mi & 0\\
                    0 & 0 & 0 & 4\mi
                \end{bmatrix},\quad\nabla_{r_2}F_{1,r_2r_3}|_{(\bm{0},\zeta)}=\begin{bmatrix}
                    0 & 0 & 0 & 0\\
                    0 & 0 & 0 & -4\mi\\
                    0 & 0 & 0 & 0\\
                    0 & -4\mi & 0 & 0
                \end{bmatrix},
        \end{equation*}
        \begin{equation*}
                \nabla_{r_3}F_{1,r_2r_4}|_{(\bm{0},\zeta)}=\begin{bmatrix}
                    4\mi\sin(\theta_4) & 0 & 0 & 0\\
                    0 & 12\mi\sin(\theta_4) & 0 & 0\\
                    0 & 0 & 20\mi\sin(\theta_4) & 0\\
                    0 & 0 & 0 & 20\mi\sin(\theta_4)\\
                \end{bmatrix},\quad
                \nabla_{r_4}F_{1,r_2r_3}|_{(\bm{0},\zeta)}=\begin{bmatrix}
                    0 & 0 & 0 & 0\\
                    0 & -16\mi\sin(\theta_4) & 0 & 0\\
                    0 & 0 & 0 & 0\\
                    0 & 0 & 0 & 16\mi\sin(\theta_4)
                \end{bmatrix},
            \end{equation*}
        \begin{equation*}
            \nabla_{\theta_4}F_{1,r_1r_3}|_{(\bm{0},\zeta)}=\begin{bmatrix}
                    0 & 0 & 0 & 0\\
                    0 & 0 & 2\mi\zeta^*\cos(2r_4) & 0\\
                    0 & 2\mi\zeta\cos(2r_4) & 0 & 2\mi r_4 s_1|s_2|\\
                    0 & 0 & 2\mi r_4 s_1|s_2| & 0\\
                \end{bmatrix}.
        \end{equation*}
        The second-order derivatives [evaluated at $(\bm{0},\zeta)$] read
        \begin{equation*}
            \begin{split}
                 \nabla_{r_1}\nabla_{r_4}F_{1,r_1r_3}|_{(\bm{0},\zeta)}&=\begin{bmatrix}
                    0 & 0 & 2\me^{-\mi\theta_4} & 0\\
                    0 & 0 & 0 & 0\\
                    -2\me^{\mi\theta_4} & 0 & 2\mi\sin(\theta_4) & 0\\
                    0 & 0 & 0 & -2\mi\sin(\theta_4)
                \end{bmatrix},\\
                \nabla_{r_1}\nabla_{\theta_4}F_{1,r_1r_3}|_{(\bm{0},\zeta)}&=\begin{bmatrix}
                    0 & 0 & -2\mi\zeta^*\cos(2r_4) & 0\\
                    0 & 0 & 0 & 0\\
                    -2\mi\zeta\cos(2r_4) & 0 & 2\mi r_4\cos(\theta_4)\cos(2r_4) & 0\\
                    0 & 0 & 0 & -2\mi r_4\cos(\theta_4)\cos(2r_4)\\
                \end{bmatrix}.
        \end{split}
        \end{equation*}
        The evaluation at the point $(\bm{0},\zeta)$ suffices to show that these are $16$ linearly-independent matrices. 
        It follows that the corresponding holonomy group $\mathrm{Hol}(A_1)$ is isomorphic to a $16$-dimensional Lie group. 
        More specifically, we have $\mathrm{Hol}(A_1)=\mathrm{U}(4)$; that is, the connection is irreducible. 
        \end{widetext}

\end{document}